\documentclass[aps,pre,twocolumn]{revtex4-1}

\usepackage{amsmath,amssymb}
\usepackage{graphicx}
\usepackage{epstopdf}

\begin{document}

\title
{
Three is much more than two 
in coarsening dynamics of cyclic competitions
}

\author{Namiko Mitarai}
\email[]{mitarai@nbi.dk}
\affiliation{
Niels Bohr Institute, University of Copenhagen,
Blegdamsvej 17, 2100 Copenhagen, Denmark. }
\author{Ivar Gunnarson}
\affiliation{
Niels Bohr Institute, University of Copenhagen,
Blegdamsvej 17, 2100 Copenhagen, Denmark. }
\author{Buster Niels Pedersen}
\affiliation{
Niels Bohr Institute, University of Copenhagen,
Blegdamsvej 17, 2100 Copenhagen, Denmark. }
\author{Christian Anker Rosiek}
\affiliation{
Niels Bohr Institute, University of Copenhagen,
Blegdamsvej 17, 2100 Copenhagen, Denmark. }
\author{Kim Sneppen}
\email[]{sneppen@nbi.dk}
\affiliation{
Niels Bohr Institute, University of Copenhagen,
Blegdamsvej 17, 2100 Copenhagen, Denmark. }


\date{\today}

\begin{abstract}
The classical game of rock-paper-scissors have inspired experiments and spatial 
model systems that address robustness of biological diversity.
In particular the game nicely illustrates that cyclic interactions allow multiple strategies to coexist for long time intervals. 
When formulated in terms of a one-dimensional cellular automata, the spatial distribution of strategies exhibits coarsening with algebraically growing domain size over time, while the two-dimensional
version allows domains to break and thereby opens for long-time coexistence. 
We here consider a quasi-one-dimensional implementation of the cyclic competition, and study the long-term dynamics as a function of rare invasions between parallel linear ecosystems. 
We find that increasing the complexity from two to three parallel subsystems allows a 
transition from complete coarsening to an active steady state where the domain size stays finite.
We further find that this transition happens irrespective of whether the update
is done in parallel for all sites simultaneously, or done randomly in sequential order.
In both cases the active state is characterized by localized bursts of 
dislocations, followed by longer periods of coarsening.
In the case of the parallel dynamics, we find that there is another phase transition between the active steady state and the coarsening state within the three-line system when the invasion rate between the subsystems is varied.
We identify the critical parameter for this transition, and show that the density of active boundaries have critical exponents that are consistent with the directed percolation universality class.
On the other hand, numerical simulations with the random sequential dynamics suggest that the system may exhibit an active steady state as long as the invasion rate is finite.
\end{abstract}


\maketitle

\section{Introduction}

Coarsening is important in a number of dynamical systems 
\cite{bray1994non,clincy2003phase}, and may be used to differentiate 
observed phenomenology into appropriated universality 
classes \cite{dornic2001critical}.
It appears in decay towards equilibrium in diverse phenomena 
as spinodal-decomposition, segregation of grains
\cite{mullins1986statistical}, opinions \cite{castellano2000nonequilibrium}, 
languages \cite{baronchelli2006topology}, populations \cite{schelling1971dynamic,vinkovic2006physical,dall2008statistical}
as well as in the ongoing tendency of biological competition 
to decrease species abundance in ecological models
\cite{karlson1984competition,chave2002comparing}.

The coarsening has been extensively studied for voter models \cite{ben1996coarsening,dornic2001critical}
and extended voter models with cyclic competition, especially 
for the 3-species cyclic competition or the rock-paper-scissors game \cite{frachebourg1996segregation,frachebourg1996spatial}. 
For the 3-species competition in one dimension, the number of 
separated populations coarsens as $t^{-3/4}$ for the random sequential dynamics where $t$ counts the number of update attempts per site.
In contrast the parallel dynamics 
provides a slower coarsening characterized by $t^{-1/2}$
\cite{frachebourg1996segregation,frachebourg1996spatial}.
One can counteract the coarsening in one dimension by introducing explicit mutation rate between species \cite{winkler2010coexistence} or by
introducing mobility \cite{venkat2010mobility}, both of which can lead to an active steady state with coexistence of all the 3 species.
Another more widely studied way is to extend it to the two-dimensional space,  
where the species domains are occasionally broken up into smaller patches, 
which in turn allow long-time coexistence of all three species \cite{szabo2007evolutionary}. 
This motivated extensive study of non-hierarchical ecosystem models as a mechanism to support coexistence of species in ecology research  \cite{karlson1984competition,bjoerlist, gilpin,Perc2007, Laird2006,Szabo2001,kerr}, and cycles are proposed to act as engines of increased diversity
in two-dimensional ecologies \cite{mathiesen2011ecosystems,mitarai2012emergence}.

In this paper, we consider cyclic predatory relations between species in a quasi-one-dimensional ecology. 
We demonstrate that when one extends a simple one-dimensional ecology to three
parallel ecologies with weak coupling between them, one obtains a hugely 
increased lifetime of all species. 
We find that this increase in lifetimes is closely
connected with on-going ``fragmentation-like" events
where invasion from one linear ecology to another initiates a positive feedback
driven by a growing divergence to the third linear ecology.
As the ecologies diverge, more successful invasions take place
between them. This opens for creation
of new patches of species within each ecology, and thereby opens for
a system where the overall invasion activity remains high.  
We quantitatively characterize the transition from the coarsening state to the active state in the parallel update case by changing invasion rate between linear subsystems, and we show that the critical behavior is consistent with the Directed Percolation (DP) universality class \cite{kinzel1983directed,hinrichsen2000non}. We further demonstrate that the random sequential update tends to make the system reach an active steady state as long as the invasion rate is finite. 

\section{Coupled linear systems of cyclic competition}
\begin{figure}[t]
\includegraphics[width=0.45\textwidth]{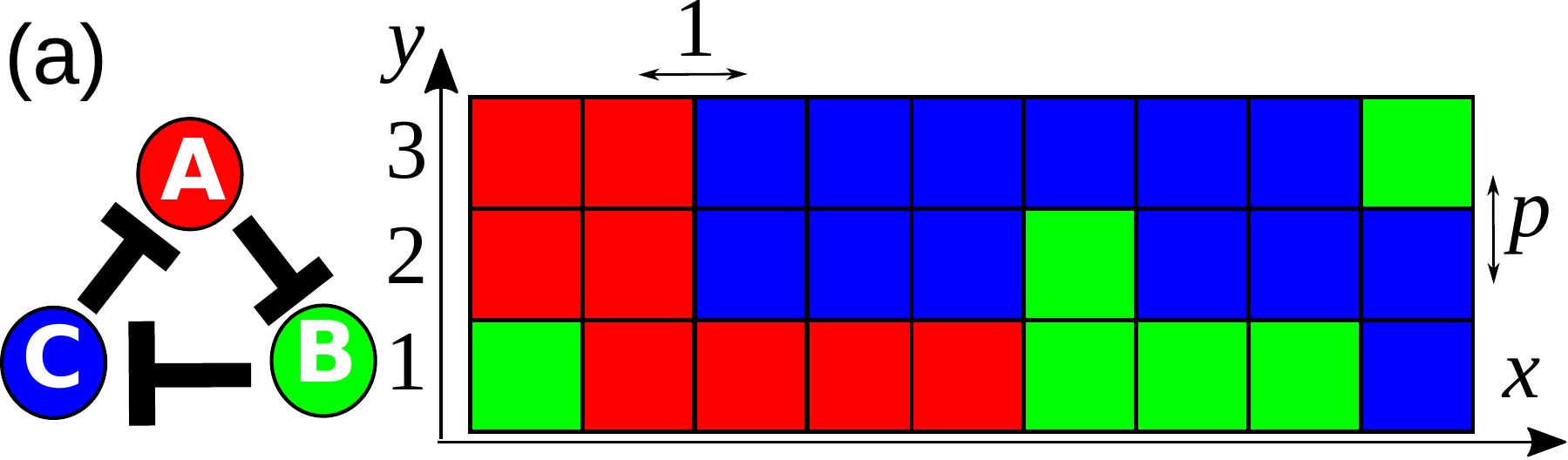}
\includegraphics[width=0.5\textwidth]{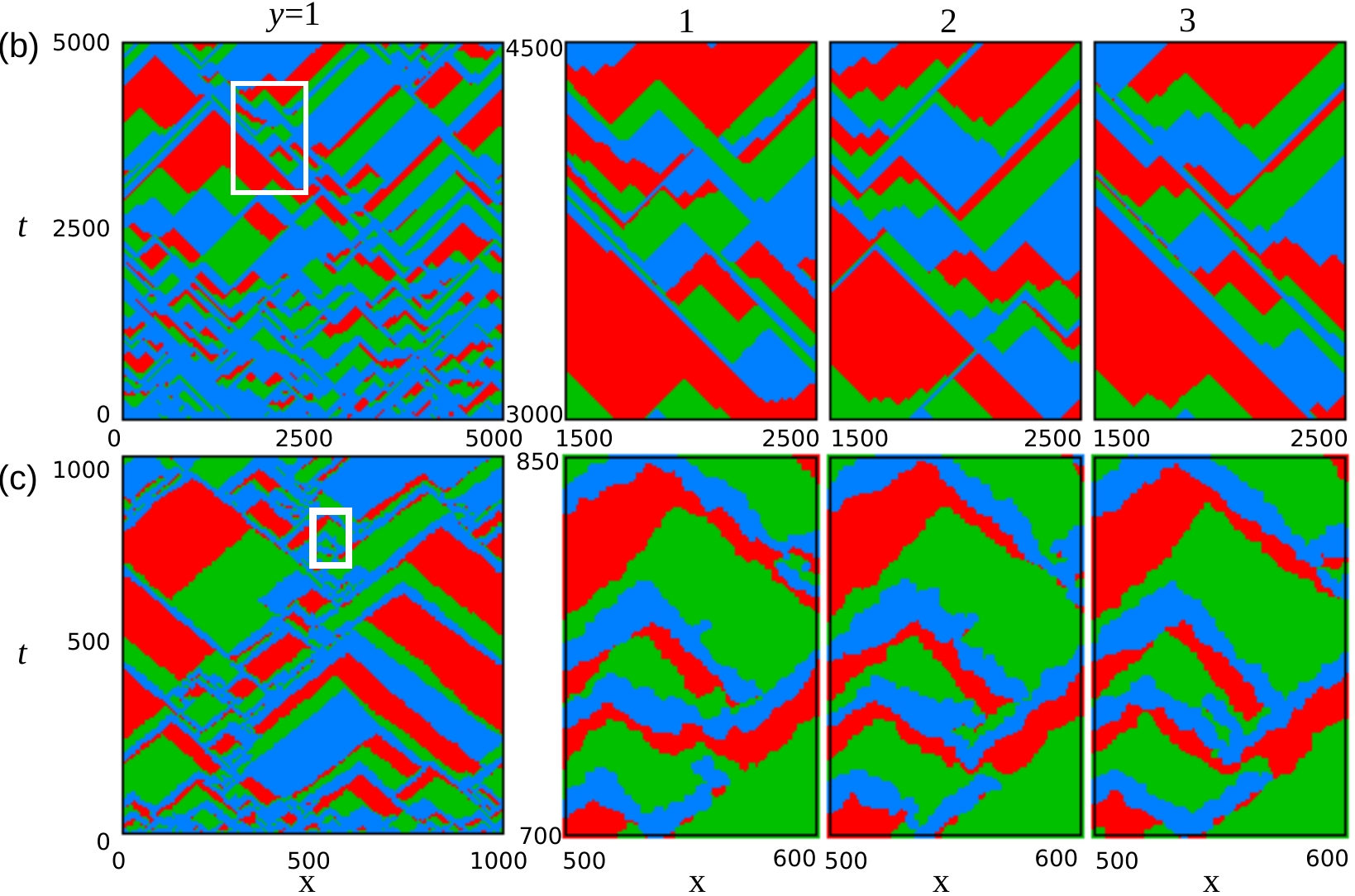}
\caption{\label{spatiotemporal} (color online)
  Model description and spatiotemporal plot with $n=3$.
(a) Schematic description of the model.
  (b) Spatiotemporal plot from the parallel dynamics with $p=0.00025$
  and $L=5000$. The left panel shows whole system until $t=5000$ for $y=1$.
  The 3 panels to the right show a magnification of part of the system
  (marked as white box in the left panel) at the corresponding locations
  for $y=1,2,3$.
  (c) As b) but for random sequential update with $p=0.05$ and $L=1000$.
} 
\end{figure}

We consider a system composed of several one-dimensional lattices that each  have a length $L$ in the $x$ direction. We in particular focus on
a stack of $n=3$ of these systems positioned on top of each other in $y$-direction
as shown in Fig.~\ref{spatiotemporal}(a). Periodic boundary conditions are imposed in both $x$ and $y$-directions.  The simulation is initialized by assigning
each lattice site to be occupied by one of the three species $A$, $B$,  or $C$ with equal probability. The species interaction is cyclic as given by the follwoing rule:
\begin{equation}
A+B\to 2A, \quad B+C\to 2B, \quad C+A\to 2C. \label{reaction}
\end{equation}
The interactions are limited to the nearest neighbor sites, and 
further limited in the vertical direction by a parameter $p$ that 
controls the vertical invasion rate relative to the 
interaction rate along the $x$ direction.  

The update can be either parallel dynamics or random sequential dynamics.

In the case of the parallel dynamics, all the bonds in the $x$ direction are updated simultaneously according to eq.~(\ref{reaction}). { For example, if
 the configuration is $ABC$, then after one update $B$ will be replaced with $A$ and $C$ will be replaced $B$ simultaneously; therefore boundaries between different species that move in the same direction will never collide.}
Then bonds in the vertical direction are selected with probability $p$ per bond (i.e., $p\cdot n \cdot L$ bonds are selected on average) and updated  { sequentially} according to eq.~(\ref{reaction}). This defines one time step in the model. 

The random sequential dynamics is defined as follows: (i) Choose a random bond in $x$ direction, and update its two neighbors according to the reactions in eq.~(\ref{reaction}).  
(ii) With a probability $p$, choose a random bond in the vertical direction,  and update its two neighbors according to the reaction in eq.~(\ref{reaction}).   One time step is here defined as $n\cdot L$ repetitions of (i) and (ii).

Irrespective of the updating rule, the system consists of domains
that each consists of populations of one of the three species.
These domains are separated by domain-boundaries that move either
left or right, as one of the populations systematically displaces 
the other.

\begin{figure}[t]
  \includegraphics[width=\hsize]{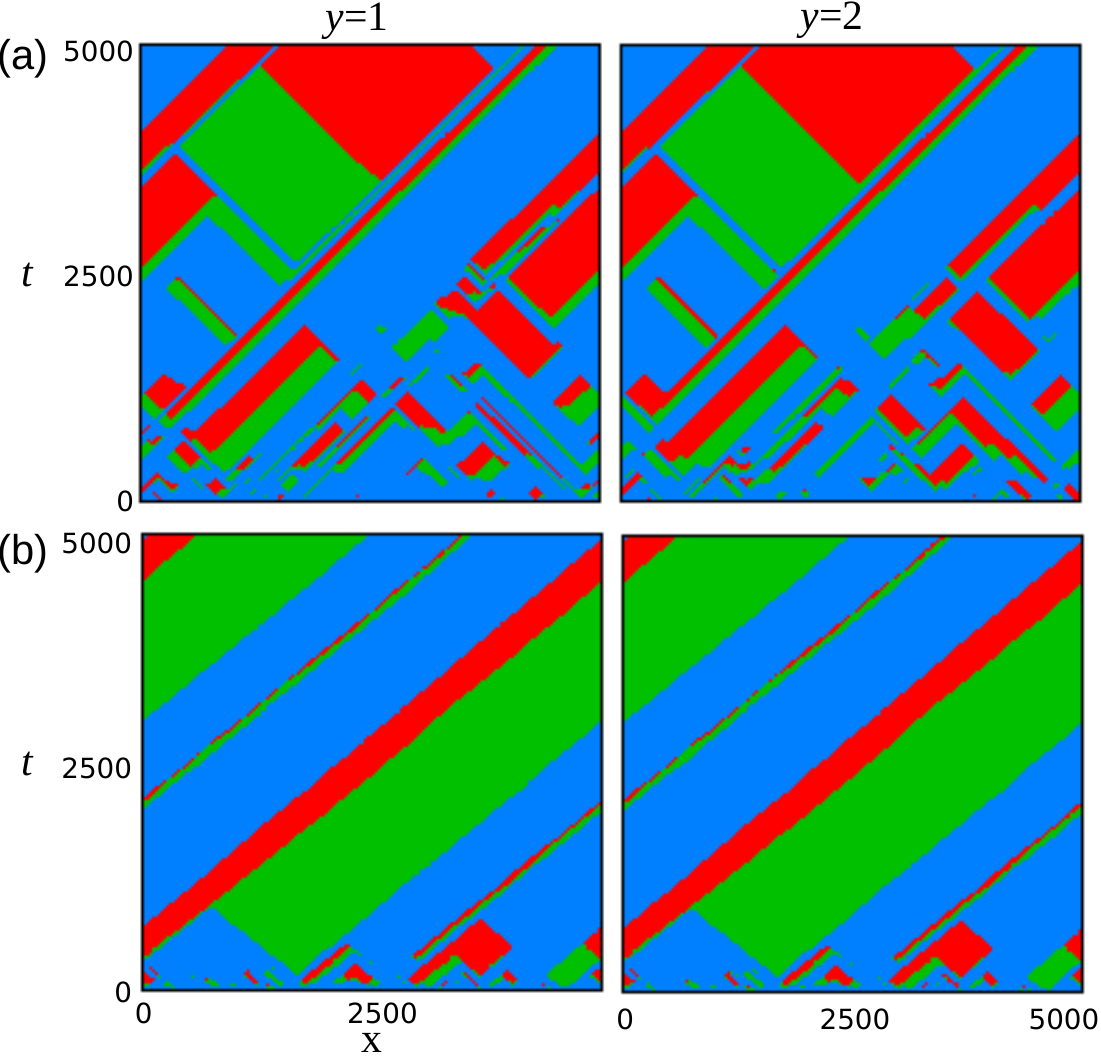}
  \includegraphics[width=\hsize]{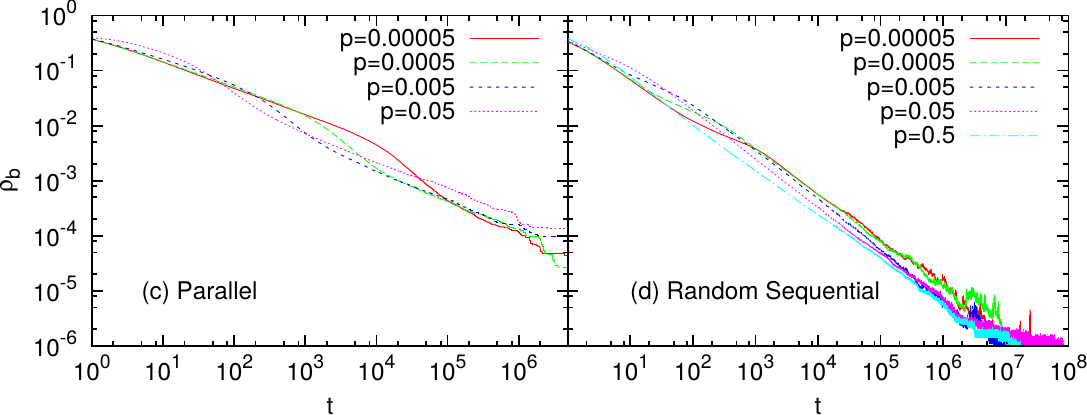}
\caption{\label{sp2lines} (color online)
  Simulation results for $n=2$.
  (a) Spatiotemporal plot for each subsystems with the parallel dynamics with $p=0.00025$  and $L=5000$. 
  (b) As (a) but for random sequential update.
  (c,d) The boundary density $\rho_b$ as a function of time for the parallel dynamics (c) and the random sequential dynamics (d) with $L=2^{24}$ for various values of $p$. 
} 
\end{figure}

In the pure one-dimensional case, i.e. $n=1$, 
coarsening happens through the collision between two moving boundaries. Such collisions eliminate the population located in the domain between the boundaries. For parallel dynamics the boundaries move at the same speed, 
and coarsening only occurs through the collision between a 
right moving boundary and a left moving one, resulting in the annihilation of both. For the random sequential update, collision of two boundaries moving in the same direction is also possible due to the fluctuating speed. 
Such a collision creates one new boundary that moves
in the opposite direction of its two parents. This makes the coarsening in the random sequential dynamics faster than that in the parallel dynamics \cite{frachebourg1996segregation}.

When parallel linear systems are added, occasional interaction between the subsystems can create new boundaries. { When $n=2$, introduction of $p$ can increase the fragmentation of the domains temporarily in early time compared to the $n=1$ case, but in the long term
  the synchronization of the two subsystems are enhanced, as shown in example spatiotemporal plots in Fig.~\ref{sp2lines}ab.
The time evolution of the density of the domain boundaries $\rho_b$,
defined as the number of domain boundaries in one of the subsystems divided by $L$,
is shown for $n=2$ for the parallel dynamics (Fig.~\ref{sp2lines}c) and
the random sequential dynamics (Fig.~\ref{sp2lines}d). We see that the coarsening continues until only the boundaries moving in parallel are left in the parallel dynamics or until only a small number (order 10) of boundaries are left in the random sequential dynamics where the noise masks the coarsening.
We could not find a value of $p$ that can stop the coarsening, neither in the parallel nor the random sequential update.
}

Interestingly, we find qualitatively different results for
$n=3$, see Figures~\ref{spatiotemporal}bc.  
Irrespective of whether one considers parallel or random sequential dynamics,
some low $p$ values make the 3-line system develop into an active steady state. In this state, the coarsening described above is balanced by 
ongoing fragmentation events that create new domains. These
events are initiated by occasional small differences between the three linear ecosystems, that subsequently cause larger divergences between
the systems. We also see this active steady state behavior 
for $n>3$ systems (data not shown), demonstrating that it 
is the transition from $n\le 2$ to $n\ge 3$ that 
fundamentally changes the overall system behavior.

\begin{figure}[t]
  \includegraphics[width=0.45\textwidth]{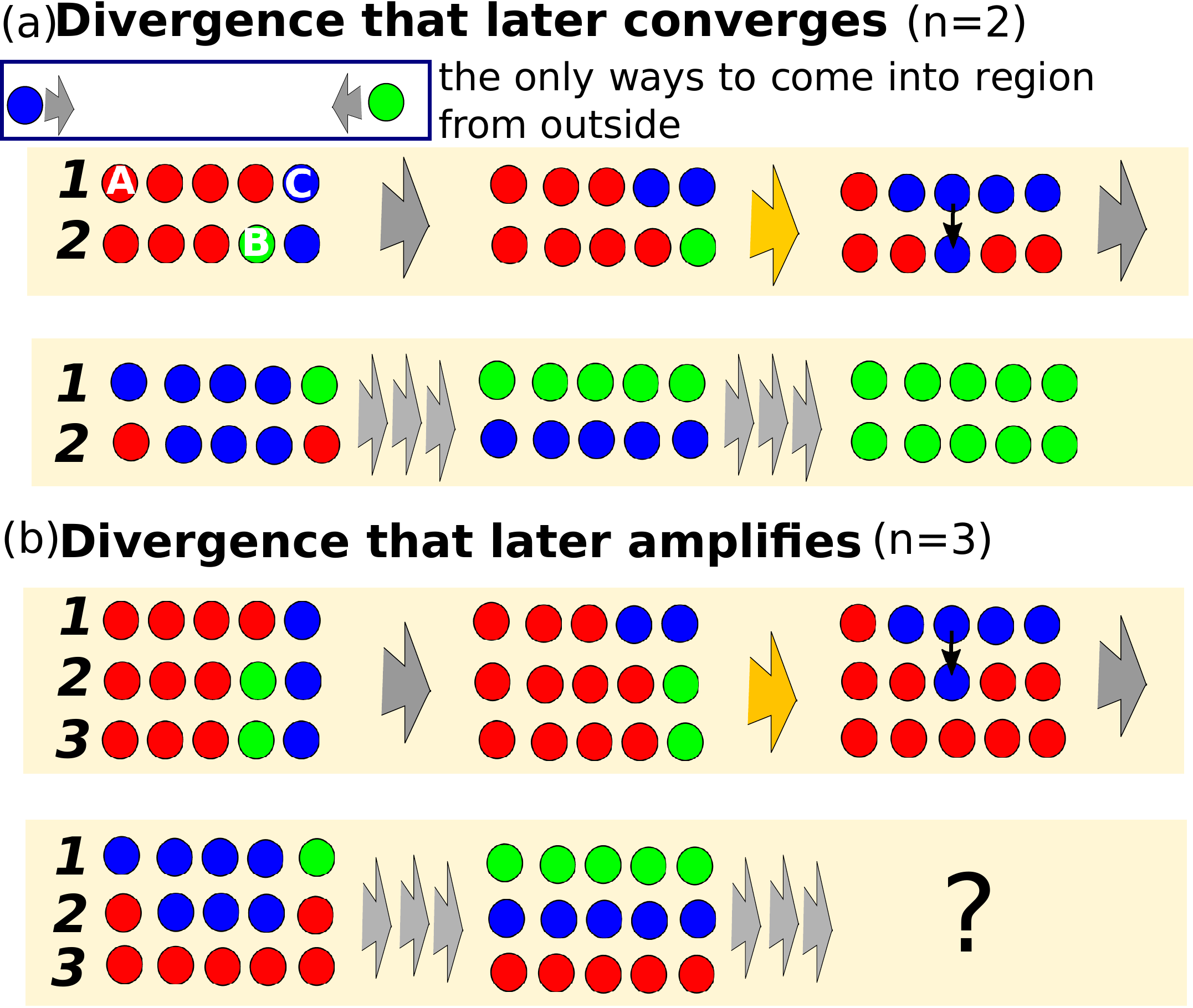}
  \caption{\label{divergence} (color online)
    Qualitative difference between $n=2$ case (a) and $n=3$ case (b).
    The vertical transfers are shown as black arrows.
    (a) In the initial situation, the difference in the considered region is only one site. Outside of the considered region, if there are
    green $B$ species to the right, they come in to the considered region in later time by eliminating the blue $C$ species. The red $A$ species cannot come in from the right, since the boundary between blue $C$ and red $A$ moves to the right. 
Similarly, from the left of the considered region, only blue $C$ can come in to the 
considered region by eliminating the red $A$ species. 
  As the time goes the species red $A$ will disappear from the considered region, allowing the green $B$ to occupy all of the considered sites in the long run.
    (b) When $n=3$, the considered region is able to keep all three species
    even starting from very similar situation as (a), opening for creation of new boundaries. 
} 
\end{figure}

{ Let us consider the difference between the $n=2$ case and the $n=3$ case.
  In order to stop coarsening, there should be possibilities for amplification of the difference between the subsystems over time.
  The only nontrivial single site difference is the situation shown in Fig.~\ref{divergence}a.
  This initial state allows temporal increase of the difference due to the diverging boundary motions.
  However, at some point a vertical invasion of a species $C$ (blue) from the subsystem 1 to
the subsystem 2 happens, making both subsystems dominated by species $C$ (blue). The only possibility to change this convergence to uniformity is the species $B$ (green) that could come in from the right side of the subsystem 1 (because it may not have species $A$ (red) that protects against invasion of $B$ (green)). But then later a vertical invasion would trigger a spread of the species $B$ (green) for both subsystems, allowing the species $B$ (green) to occupy all of the considered sites. Therefore one cannot keep the difference between the subsystems with $n=2$, and the system will always tend to coasen over long time.  
  
On the contrary, a similar situation for $n=3$ case 
can keep the difference among the subsystems.
As shown in Fig.~\ref{divergence}b, the additional third subsystem can keep
species $A$ (red) in the considered region, and the configuration in which all the three species present in the considered region enables for various ways of creating new boundaries.
The ability to keep all the species in the same region is needed to keep the active steady state. 
}

The transition to the active steady state in the $n=3$ case is quantitatively different between the parallel and random sequential update. Note that  Fig.~\ref{spatiotemporal}b for parallel dynamics shows the result with $p=0.00025$ with $L=5000$, while  Fig.~\ref{spatiotemporal}c for random sequential dynamics shows the result with $p=0.05$ with $L=1000$. 
We also find some qualitative difference in the transition between the coarsening state and the active state when varying $p$.
In the subsequent sections, we further quantify the coarsening dynamics for the two updating rules. 

\section{Parallel dynamics}

\begin{figure}[t]
  \includegraphics[width=0.8\hsize]{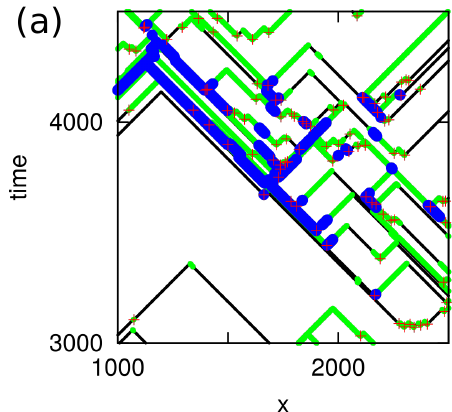}
  \includegraphics[width=0.95\hsize]{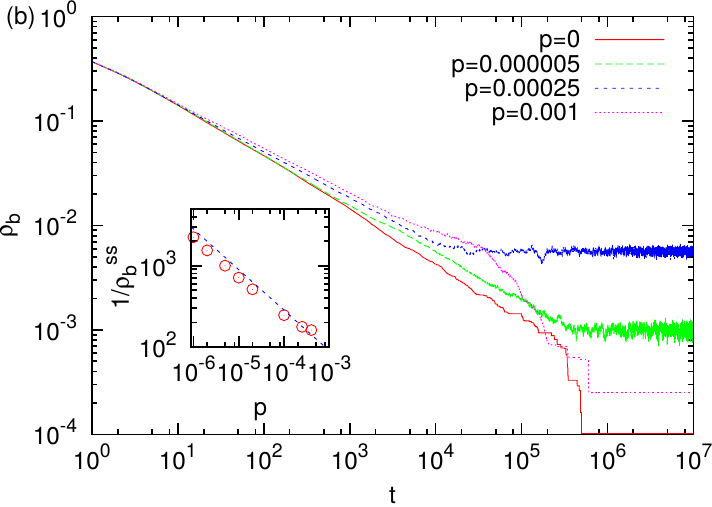}
  \caption{\label{div-p}  (color online)
        (a) Development of boundaries using 
        with invasion probability $p=0.00025$.
    { Thin black symbols mark the points where the boundary sites have 
    same species in all 3 subsystems. Thick green circles mark boundary sites where one subsystem deviates, whereas thick blue circles mark boundary sites where all subsystems carry different species. The red crosses show events where species from subsystem 2 or 3 invade the subsystem 1. }
    (b) Density of boundaries per subsystem $\rho_{b}$ simulated
    with system size $L=2^{20}$ and $n=3$.
    The $p=0$ case is equivalent to the $n=1$ case with long-time coarsening as $t^{-1/2}$.
    Inset: The average domain size in the active steady state $1/\rho_{b}^{ss}$ vs. $p$. The dashed line shows $2\sqrt{2}/\sqrt{p}$ (see text).
  }  
\end{figure}

\begin{figure}[t]
\includegraphics[width=\hsize]{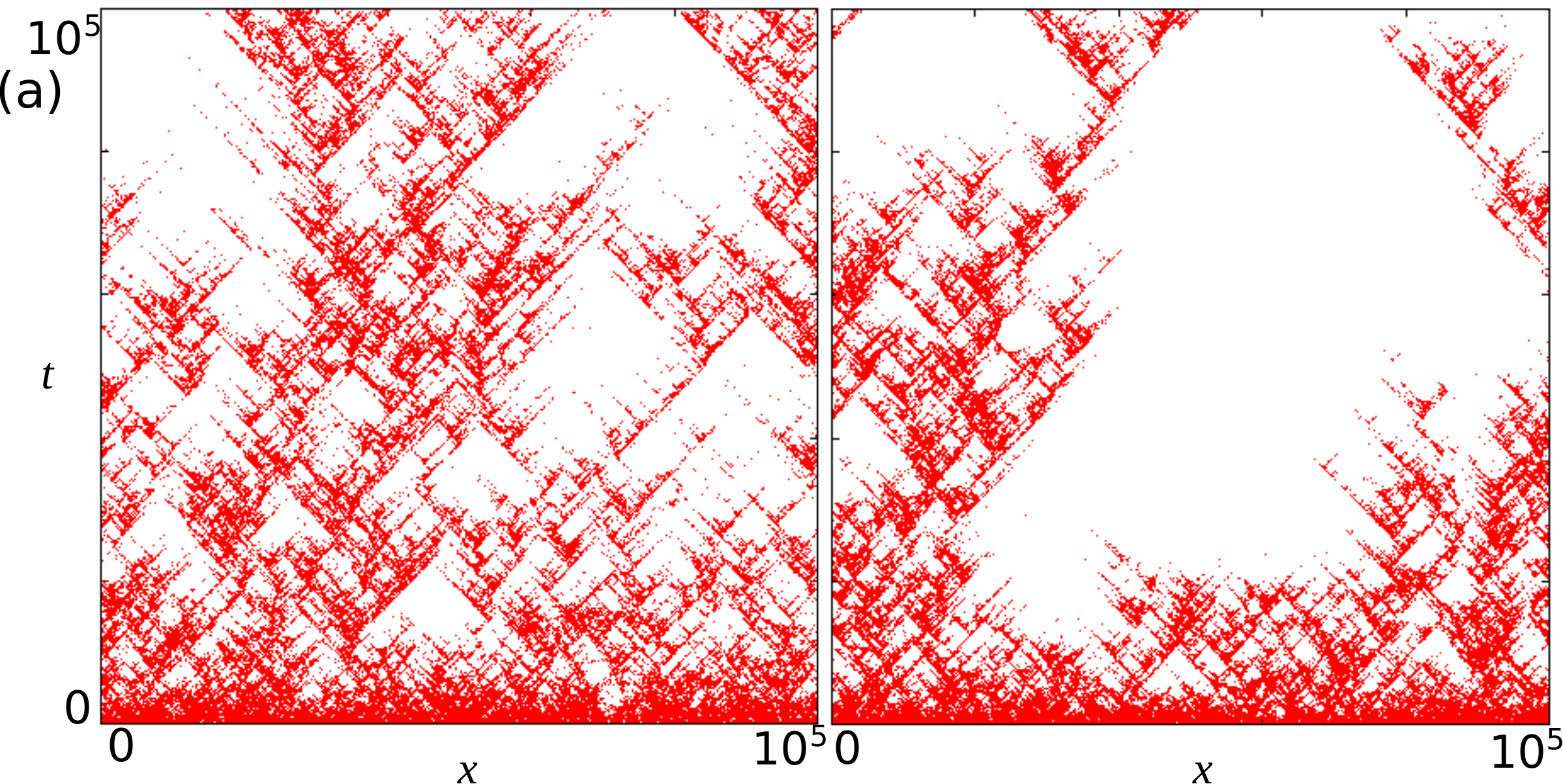}
\includegraphics[width=.8\hsize]{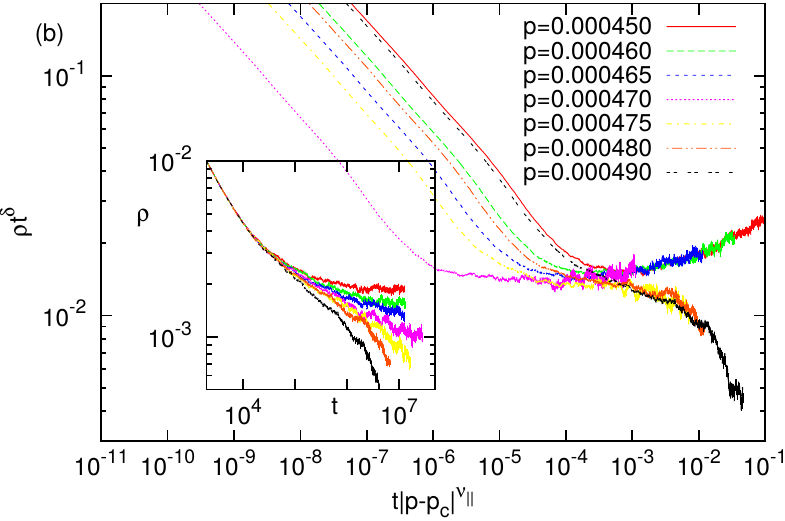}
\includegraphics[width=.8\hsize]{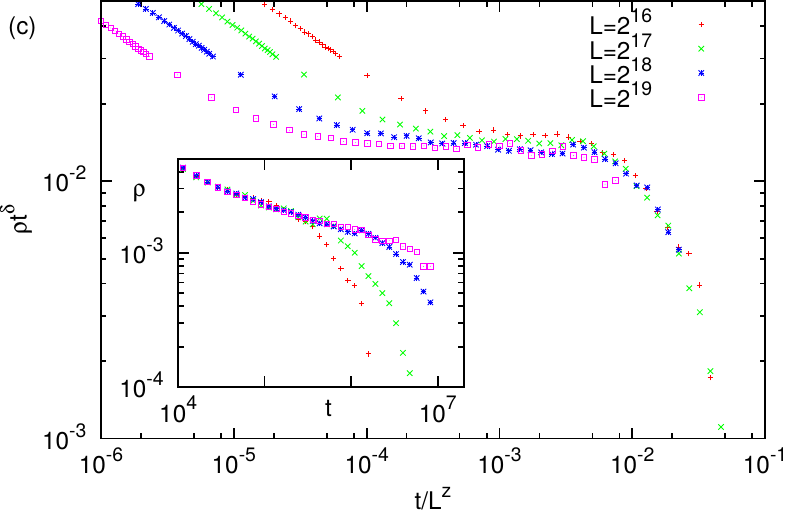}
  \caption{\label{vertical} (color online)
(a) Large-scale behavior of the vertical transfer events for
parallel update.
  The red dots show the invasions of species to subsystem 1 from subsystem 2
  or 3.
  The figure shows a part of a $L=1000000$ system.
The left panel is for $p=0.00025$ and the right panel is for $p=0.0005$.   
(b) Scaling plot for the simulations of a system size $L=2^{26}$ 
using $p$ values close to the critical invasion rate $p_c$. 
We fitted this to be $p_c=0.000471$ and used 1+1 dimensional DP scaling exponents $\delta=0.0159$ and $\nu_{||}=1.733$ for rescaling. The
inset shows the time evolution of $\rho$ without such rescaling.
(c) Finite size scaling plot at $p=0.000475$ with $z=1.581$. The inset shows time evolution of $\rho$ without rescaling. 
  }
\end{figure}
In the case of the parallel dynamics there is no noise in the 
horizontal movement of the boundaries. Thus if we initially synchronize all the three subsystems (i.e, $S_i^{1}=S_i^{2}=S_i^{3}$ for all $i$, with $S_i^k$ being the species name at the site with $x=i$ and $y=k$), then the subsystems will stay synchronized and the dynamics will be identical to the 
one-dimensional system, irrespective of the value of $p$.
Therefore, to maintain the active steady state there must remain 
differences among the three subsystems{ , where a case study was already shown in Fig.~\ref{divergence}b.}

{ Figure~\ref{div-p}a shows the motion of the boundaries ($S_i^{k}\ne S_{i+1}^k$)  from the simulation shown in Fig.~\ref{spatiotemporal}b. The boundaries between the domains in the subsystem 1 are shown. The green symbols mark boundary sites where one of the subsystems is different whereas thick blue symbols mark sites where all subsystems differ. The red crosses show the successful invasions of the species to line 1 from the subsystem 2 or 3. We observe that such vertical transfers create new interfaces.}

The competition between the gradual alignment and the creation of new boundaries by vertical invasion causes a transition in behavior between a low $p$-case where there is a sustained activity, to a high $p$-case where
the system persistently coarsen.
Figure ~\ref{div-p}b shows the development in the boundary density 
$\rho_{b}$ 
for different  values of $p$ in a large system. 
When $p$ is increased from $p=0.000005$ to $p=0.00025$, we see 
that the system settles in a steady state with constant number of domains (boundaries). Also we see that an increased $p$ increases the number
of such boundaries. However, further increase to $p=0.001$ shows the collapse of the active steady state to the coarsening mode that is also found for 
the case of isolated subsystems ($p=0$). This is because high $p$ makes 
gradual alignment happen too often compared to the creation of new boundaries. Then in the high $p$ case all the subsystems act as the same in the long term, which is equivalent to the 
one-dimensional case.

Considering the system in an active steady state,
we can estimate the average time $\Delta t$ it takes between
the first creation of divergent boundaries at time $t^*$ (an event like the vertical transfer in Fig.~\ref{divergence}b), to the first transfer of divergent states among the 3 subsystems.
This time is given by the transfer rate $p$ per site 
in a linearly growing divergent region between two subsystems.
Assuming that the event occurs when the cumulative probability is one, we expect
\begin{eqnarray}
\nonumber
\int_{t^*}^{t^*+\Delta t} p \cdot t \cdot {\rm d}t & \approx & 1,
\end{eqnarray}
which gives $\Delta t  \approx  \sqrt{2/p}$.
As the boundary motion is ballistic, a new transfer occurs between a divergent region of size
$\Delta {\ell} \approx 2\Delta t \approx 2\sqrt{2/p}$.
Or said in another way, then the average steady state domain size
for small $p$ should scale as $\Delta {\ell}$, and one indeed sees it in the insert of 
Fig.~\ref{div-p}b.

The spatiotemporal plots of successful vertical invasions are shown in
Fig.~\ref{vertical}a for { $p=0.00025$ (left) and $p= 0.0005$ (right)}.
One observes ballistic lines that follows the propagating boundaries until they occasionally disappear due to local synchronization among the
three subsystems. Also one observes the occasional creation of new boundaries from
old ones. These creation events seem to occur 
as dense bursts of activity.
The large-scale structure of this spreading birth and death process
reminds us of the directed percolation (DP) class of models in 
1+1 dimension \cite{hinrichsen2000non}.

For small $p$ (Fig.~\ref{vertical}a left) the alignment is so slow that vertical transfer dominates and activity is sustained. 
At large $p$ values, the faster alignment between the three sub-systems 
makes it more difficult to maintain sufficient divergence to sustain
on-going vertical transfer events (Fig.~\ref{vertical}a right),
and ultimately the whole system align to form a few parallel moving domains. It should be noted that the smallest ``unit'' of this DP-like structure is not one site but instead given by the length and time scale of $1/p\sim 10^3$. 

We conjecture that the transition from the active state to the coarsening state belongs to the DP universality class.
Note that the absorbing state at large $p$ is the state where all three subsystems are synchronized, still leaving possibility for some 
diversity with some moving but synchronized boundaries.
We, therefore, chose to study the density of the {\sl active} sites $\rho$, defined as the density of the boundaries that contains sites which are not completely aligned with other subsystems.
The development of $\rho$ in inset of Fig.~\ref{vertical}b 
illustrates transient coarsening up to  $t\sim 10^4$, 
after which it changes to either a steady state density 
or collapses to zero density.
We rescaled these data by using 1+1 dimensional DP exponents \cite{jensen1999low,hinrichsen2000non} $\delta=\beta/\nu_{\parallel}=0.0159$ and $\nu_{\parallel}=1.733$.  By fitting { the critical $p$ at the transition, $p_c$,} to $0.000471$, we obtain a data collapse that is consistent with the DP universality class, except for the initial transient regime (Fig.~\ref{vertical}b). Figure~\ref{vertical}c shows the finite size scaling using another DP-scaling exponent $z=\nu_{\parallel}/\nu_{\perp}=1.581$. Thus both the time coarsening and the finite size dependence are consistent with the DP-universality class.

\section{Random sequential dynamics}
The behavior of the random sequential dynamics is more complicated, because three subsystems can spontaneously de-synchronize due to the randomness of the movement of the boundaries in the respective sub-systems. For example, if all the three subsystems are locally identical with two left moving interfaces $ABC$ each, it is possible that the two boundaries in the subsystem 1 merge spontaneously to make $AAC$ at next time step. This suddenly creates one right moving interface, very similar to the situation in Fig.~\ref{divergence}b, allowing further diversification. Collapse of interfaces that are moving in the same direction is the reason why the random sequential dynamics coarsen faster than the parallel dynamics in one-dimensional system. With vertical coupling it however also provides an additional way to create more boundaries. 

\begin{figure}[t]
  \includegraphics[width=\hsize]{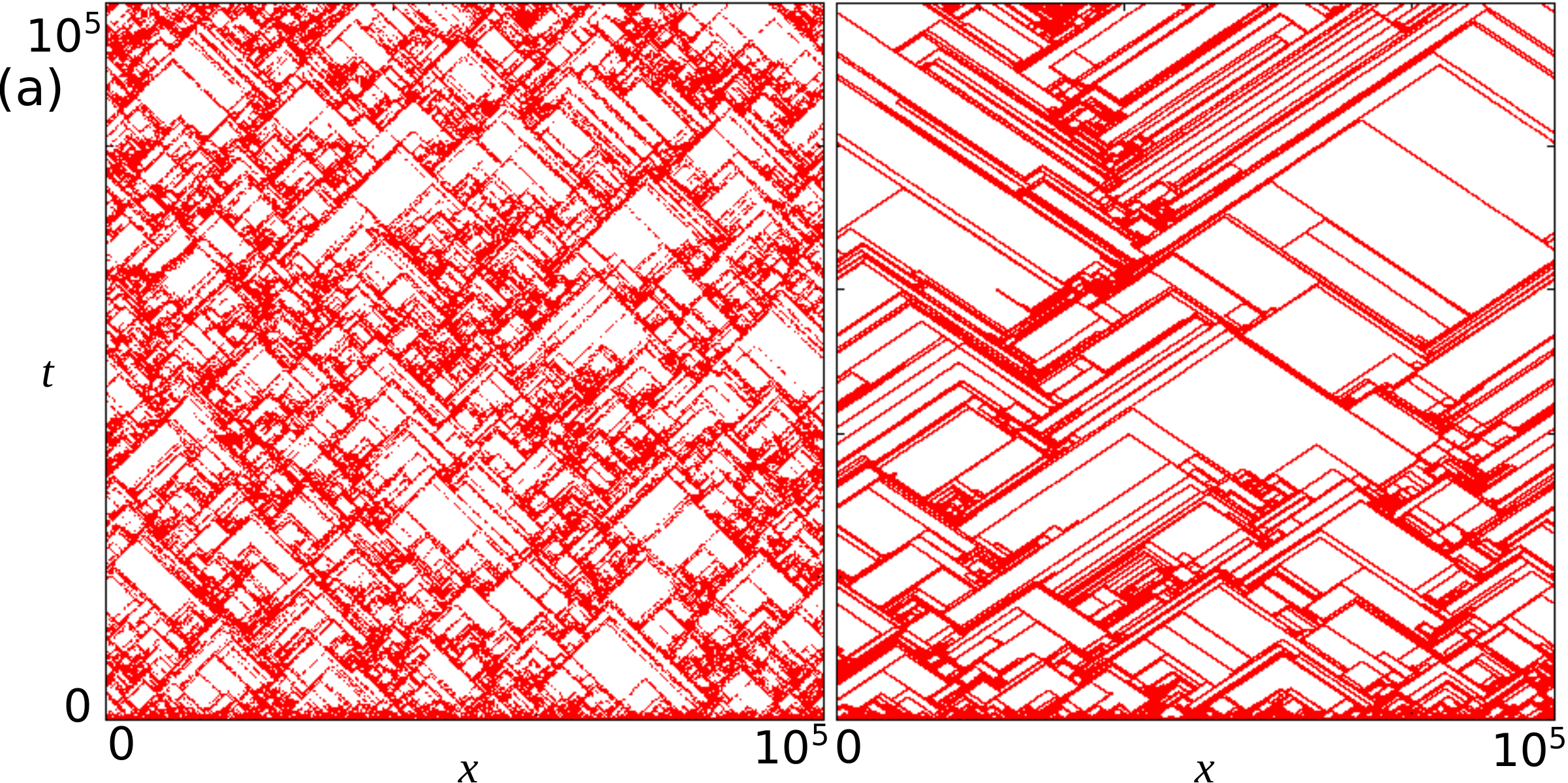}
\includegraphics[width=0.9\hsize]{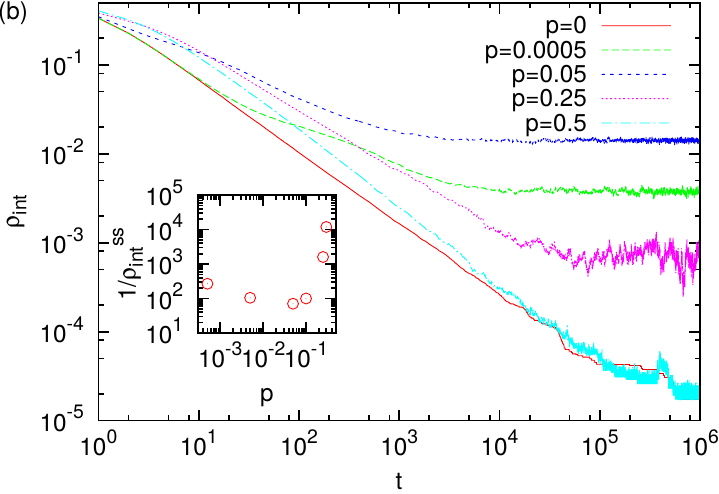}
\caption{\label{interface2to20r} (color online)
  (a)  Large scale structure of the vertical transfer events for $L=100000$ with random sequential update.
  The red dots shows the invasion of species to the subsystem 1 from the subsystems 2 and 3.
  The left panel shows $p=0.0005$, and the right panel shows $p=0.25$.
  (b) Density of interface per subsystem with $L=2^{20}$ and $n=3$ The random sequential update for for $p=0$ (equivalent to $n=1$,  with long-time behavior $t^{-3/4}$), $p=0.0005,0.05,0.25,$ and $0.5$. Inset shows the average domain size in the steady state $1/\rho_{b}^{ss}$ as a function of $p$.
The data for $p=0.3$ was obtained from the simulation with $L=2^{24}$.
} 
\end{figure}

First of all, this allows the three-line system to maintain 
an active steady state for much higher $p$-values than in 
the parallel update case,
a robustness that reflects the more frequent diversification events. 
Furthermore, the fully synchronized state is no longer an absorbing state. Numerically this seems to result in loss of a clear transition with changing
value of $p$. This manifests itself in the spatiotemporal plot of the successful invasion to the subsystem 1 from other subsystems shown  in Fig.~\ref{interface2to20r}a. In contrast to the parallel dynamics case in Fig.~\ref{vertical}b, we see continuous ballistic trajectories that closely follow the
moving domain boundaries in one of the sub-systems. 
{ This is because the random fluctuations of the boundary motions keep desynchronizing the subsystems to allow vertical transfers, until the boundary disappears. In other words, 
in}
the random update case, an active site
can only be annihilated by meeting another active site.
This in itself is qualitatively different from the DP-universality class.

The time evolution of the boundary density $\rho_{b}$ is shown in Fig.~\ref{interface2to20r}b. The $p=0$ (pure one-dimensional) case shows coarsening that declines as $t^{-3/4}$ in the long time limit. 
Introducing a small finite $p$ increases $\rho_{b}$ compared to
the $p=0$ case. Increasing $p$ to $0.0005$ and further to $0.05$ allowed
the system to reach a steady state with a constant $\rho_b$.
Further increase of $p$ decreased $\rho_b$ at the steady state, but no sudden collapse was observed. 
The corresponding non-monotonous behavior of the average steady state domain size $1/\rho_{b}^{ss}$ as a function of $p$ is shown in inset. 
Simulations with larger $p$ always allowed us to find correspondingly 
larger systems with an active steady state within the range that we could test numerically (we tried up to size $L=2^{26}$).

\section{Discussion}
We have shown that the transition from coarsening of domains of rock-paper-scissors game to an active steady state with a finite level of interface density requires at least three coupled linear subsystems. 
With such systems, it becomes difficult to synchronize all three, which in turn gives rise to the creation of new domains through the mutual invasions.

With the parallel dynamics, the complete synchronization of the three linear subsystems acts as an absorbing state, and the system exhibits a transition from the active steady state where subsystems never synchronize to the absorbing state. We have shown that the transition is consistent with the DP universality class in 1+1 dimension.

It has been conjectured that the short-range process is a 
requirement for the DP universality class \cite{hinrichsen2000non}. The observation of DP class in the present model was unexpected because of the apparent long-range correlation between ballistic boundaries. When subsystems are coupled by rare invasions, however, the invasion from other subsystems breaks up this correlation, and the interaction between domains becomes ``short range'' when viewed on length scales larger than $1/p$. Since the critical $p$ happens to be about $0.0005$, the DP behavior appears only after long transient in large systems.

When the update is random and sequential, the synchronized state is no longer an absorbing state. It is then possible that the active steady state may exist as long as $p$ is finite. Since $p$ is the rate per site for the vertical invasion, we can in principle consider $p\to \infty$ limit, where all the three subsystems stay synchronized. Note that this limit is not exactly the same as the pure one-dimensional system,  since if a boundary of one of the three subsystems proceeds more than average by chance, that will be copied to other subsystems immediately, namely fluctuation tends to make the interface motion slightly faster, which may result in faster coarsening than one-dimensional system. We did not identify any transition in the systems behavior at finite $p$ in the random sequential dynamics.
We speculate that this type of systems deviates fundamentally from the DP class because the active { (desynchronized)} boundary cannot "die" by itself, but rather needs another active site to be eliminated. { The feature that there is no spontaneous death process differenciates it from the DP process.}

Overall lesson from this work is that three is much more than two and
provides an engine for sustained yet dynamic heterogeneity 
in spatial rock-paper-scissors game.
Thus parallel systems open for  a qualitatively different way of sustaining patchiness  from increasing  the number of species in a cycle of invasions \cite{frachebourg1996segregation,frachebourg1996spatial}. 

\begin{acknowledgments}
  This work was supported by the Danish National Research Foundation. 
  NM is grateful to M. H. Jensen for fruitful discussions. 
\end{acknowledgments}
\bibliography{triple-refs,plants}
\end{document}